\documentclass{llncs}
\usepackage{times}

\usepackage[utf8x]{inputenc}
\usepackage[UKenglish]{babel}

\usepackage{amssymb}
\usepackage{amsmath}

\usepackage{xspace}   

\usepackage{tikz}
\usetikzlibrary{arrows,automata,positioning}
\usepackage[unicode]{hyperref}

\newcommand{\nb}[1]{}

\begin{document}
\title{A SAT Attack on the Erd\H{o}s Discrepancy Conjecture}
\date{}

\author{Boris Konev \and Alexei Lisitsa}
\institute{
Department of Computer Science\\
University of Liverpool,
United Kingdom\\
}

\maketitle
\begin{abstract}

In 1930s Paul Erd\H{o}s conjectured that for any positive integer $C$ in any
infinite $\pm 1$ sequence $(x_n)$ there exists a subsequence $x_d, x_{2d},
x_{3d},\dots, x_{kd}$, for some positive integers $k$ and $d$, such that
$\mid \sum_{i=1}^k x_{id} \mid >C$.  
The conjecture has been referred to  as one of the major open problems in 
combinatorial number theory and discrepancy theory.  
For the particular case of $C=1$ a human proof of the conjecture exists; for
$C=2$ a bespoke computer program had generated sequences of length $1124$
of discrepancy $2$,
but the status of the conjecture remained open even for such a small bound.
We show that by encoding the problem into Boolean satisfiability and applying the 
state of the art SAT solver, one can obtain a discrepancy $2$ sequence of length $1160$
and a \emph{proof} of the Erd\H{o}s discrepancy conjecture for
$C=2$, claiming that no discrepancy 2 sequence of length $1161$, or more, exists.    
We also present our partial results for the case of $C=3$.
\end{abstract}

\section{Introduction} 
Discrepancy theory is a branch of mathematics dealing with 
irregularities of distributions of points in some space in combinatorial, measure-theoretic
and geometric settings~\cite{Beck,Chazelle,Matusek,BeckSos}. 
The paradigmatic combinatorial discrepancy theory  setting can be described in
terms of a hypergraph $\mathcal{H} = (U,S)$, that is, a set $U$  and a family
of its subsets $S \subseteq 2^{U}$. 
Consider a colouring $c : U \rightarrow \{+1,-1\}$ of the elements of $U$ in  
\emph{blue} $(+1)$ and \emph{red}  ($-1$) colours. Then one may ask whether there 
exists a colouring of the elements of ${U}$ such that in every element of $S$ colours are
distributed uniformly or a discrepancy of colours is always inevitable.
Formally, the discrepancy (deviation from a uniform distribution) of a hypergraph $\mathcal{H}$ is defined as
$\min_{c} (\max_{s \in S} \,| \sum_{e \in s} c(e) |\,)$.
Discrepancy theory also has practical applications 
in {computational complexity}~\cite{Chazelle},
complexity of communication~\cite{DBLP:journals/dam/Alon92}
and {differential privacy}~\cite{Muthukrishnan:2012:OPH:2213977.2214090}. 

One of the oldest problems of discrepancy theory is the 
discrepancy of hypergraphs over the set of natural numbers with the  subsets
(hyperedges) forming arithmetical progressions over this set~\cite{MS96}. 
 Roth's theorem~\cite{Roth64}, one of the main results in the area, states that
for the hypergraph formed by the arithmetic progressions in $\{1,\dots, l\}$, that is
$\mathcal{H}_l=(U_l, S_l)$, where $U_l = \{1,2, \ldots, l\}$ and
elements of $S_l$ being of the form $(ai+b)$ for arbitrary $a, b$,
the discrepancy grows at least as $\frac{1}{20}l^{1/4}$.

Surprisingly, for the more restricted case of \emph{homogeneous} arithmetic progressions of the form
$(ai)$, the question of the discrepancy bounds is open for more than eighty years.   
In 1930s Paul Erd\H{o}s conjectured~\cite{Unsolved} that
the discrepancy is unbounded.
This conjecture became known as the Erd\H{o}s 
discrepancy problem (EDP) 
and its proving or  disproving has been referred to
as one of the major open problems in combinatorial number theory and discrepancy theory~\cite{Beck,BeckSos,NiTa}.

The problem can be naturally described in
terms of sequences of $+1$ and $-1$ (and this is how Erd\H{o}s himself introduced it).
Then Erd\H{o}s's conjecture states that for any $C>0$ in any
infinite $\pm 1$ sequence $(x_n)$ there exists a subsequence $x_d, x_{2d},
x_{3d},\dots, x_{kd}$, for some positive integers $k$ and $d$, such that
$\mid \sum_{i=1}^k x_{id} \mid >C$. 
The general definition of discrepancy given above can be specialised as follows.
The discrepancy of 
a finite  $\pm 1$ sequence $\bar{x} = x_1,\dots,x_l$ of length $l$
can be defined as 
$\max_{d=1 ,\ldots, l} (\mid \sum_{i=1}^{\lfloor \frac{l}{d} \rfloor } x_{i d} \mid)$. For an infinite sequence
$(x_n)$ its discrepancy is the supremum of discrepancies of all its initial
finite fragments. 

For random $\pm 1$ sequences of length $l$ the discrepancy grows as
$l^{1/2+o(1)}$ and the explicit constructions of a sequence with slowly growing
discrepancy at the rate of $\log_3 l$ have been demonstrated~\cite{gowers,BCC10}. It
is known~\cite{Mathias}
that discrepancy of any infinite $\pm1$ sequence can not be bounded by 1, that
is, Erd\H{o}s's conjecture holds for the particular case $C=1$. For all other
values of $C$  the status of the conjecture remained unknown.  Although widely
believed not to be the case, there was still a possibility that an infinite
sequence of discrepancy 2 existed.

The EDP has attracted renewed interest in 2009-2010  as it became a topic of the
Poly\-math project \cite{Polymath}, a widely publicised endeavour in
collective math initiated by T.~Gowers~\cite{gowersblog}.   As part of this
activity (see discussion in \cite{Polymath}) an attempt has been made
to attack the problem using computers. A purposely written computer 
program had successfully found $\pm1$ sequences of length 1124 having
discrepancy 2; however, it failed to produce a discrepancy $2$ sequence of a larger
length and it has been claimed that ``given how long a finite sequence can
be, it seems unlikely that we could answer this question just by a clever
search of all possibilities on a computer''~\cite{Polymath}.

In this paper we settle the status of the EDP for $C=2$. We show that by
encoding the problem into Boolean satisfiability and applying the state of the
art SAT solvers, one can obtain a sequence of length $1160$ of discrepancy 2
and a proof of the Erd\H{o}s discrepancy conjecture for $C=2$, claiming that no
sequence of length $1161$ and discrepancy $2$ exists.  We also present our
partial results for the case of $C=3$ and demonstrate the existence of
a sequence of length $13\,000$ of discrepancy $3$.  

\section{SAT Encoding}
Checking that a $\pm 1$ sequence of length $l$ has discrepancy $C$ 
is quite straightforward and so for the existence claims the specific encoding details are
of limited interest and could be left as an exercise to the reader. The negative 
results (that is, our claim that no infinite discrepancy $2$ sequence exists),
however, require us to give a short description of our SAT encoding of the EDP.
The encoding in full for all cases discussed in this paper and the program generating 
the encoding of the EDP for arbitrary given values of $C$ and $l$ can be found in~\cite{KL14}.

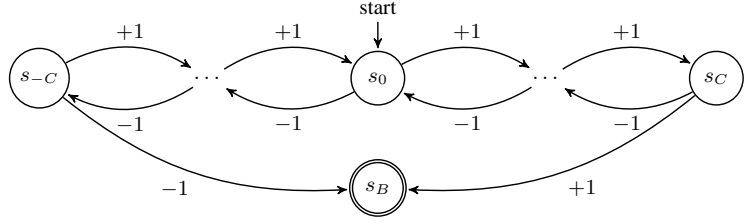
\begin{figure}[t]
\begin{center}
\scalebox{0.9}{
\begin{tikzpicture}[->,>=stealth',shorten >=1pt,auto,node distance=2.5cm, semithick]
\node[initial above,state] (Z)              {$s_0$};
\node                (O)  [right of=Z] {$\dots$};
\node[state]         (C)  [right of=O] {$s_C$};
\node                (Om) [left of=Z]  {$\dots$};
\node[state]         (Cm) [left of=Om] {$s_{-C}$};
\node[state, accepting]         (B) [below=.8cm of Z] {$s_B$};
\path 
      (Z) edge [bend left] node {$+1$} (O)
      (O) edge [bend left] node {$-1$} (Z)
      (O) edge [bend left] node {$+1$} (C)
      (C) edge [bend left] node {$-1$} (O)
      (C) edge [bend left=21,below right] node {$+1$} (B)
      (Z) edge [bend left] node {$-1$} (Om)
      (Om) edge [bend left] node {$+1$} (Z)
      (Om) edge [bend left] node {$-1$} (Cm)
      (Cm) edge [bend left] node {$+1$} (Om)
      (Cm) edge [bend right=21,below left] node {$-1$} (B)
;
\end{tikzpicture}
}
\end{center}
\caption{Automaton $\mathcal{A}_{C}$.\label{fig:automaton}}
\end{figure}

We employ the automata based approach similar to the encoding of temporal 
formulae for bounded model checking~\cite{BMC}.
In Figure~\ref{fig:automaton} we give an automaton that accepts a $\pm 1$ 
word of length $m$ if, and only if, the word represents a $\pm1$ sequence
$y_1,\dots,y_m$ such that $\sum_{i=1}^m\mid y_i\mid > C$ (and for all $m'<m$
it holds $\sum_{i=1}^m\mid y_i\mid \leq C$). 
Notice that if a subsequence $x_d,x_{2d},\dots, x_{kd}$ of 
$\bar x = x_1,\dots, x_l$ contains less than $C$ elements, this subsequence
does not contribute to the discrepancy of $\bar x$.
It should be clear then that if for every $d:1\leq d \leq \lfloor \frac{l}{C+1} \rfloor$
the automaton $\mathcal{A}_C$ \emph{does not} accept the subsequence
$x_d, x_{2d},\dots, x_{kd}$, where 
$k = \lfloor \frac{l}{d}\rfloor$\nb{please check \emph{all} values in this paragraph!}, then the discrepancy of the sequence $\bar x$
does not exceed $C$.

The trace of the automaton $\mathcal{A}_C$ on the subsequence $x_d,
x_{2d},\dots, x_{kd}$ can be encoded by a Boolean formula in the obvious way.
To explain representation details, first consider
\begin{equation}\label{eq:1}
\begin{array}{ll}
\phi_{(l,C,d)} = s_{0}^{(1,d)} \bigwedge\limits_{i=1}^{\lfloor \frac{l}{d}\rfloor}\bigg[
  & \bigwedge\limits_{-C\leq j < C} \left(s_{j}^{(i,d)}\land p_{id} \rightarrow s_{j+1}^{(i+1,d)}\right) \land \\
  & \bigwedge\limits_{-C< j \leq C} \left(s_{j}^{(i,d)}\land \lnot p_{id} \rightarrow s_{j-1}^{(i+1,d)}\right) \land \\
  & \left(s_{C}^{(i,d)}\land p_{id} \rightarrow B\right) \land \\
  & \left(s_{-C}^{(i,d)}\land \lnot p_{id} \rightarrow B\right)\bigg],
\end{array}
\end{equation}
where the intended meaning is that 
proposition $s_{j}^{(i,d)}$ is true if, and only if, the automaton $\mathcal{A}_C$
is in the state $s_j$ having read first $(i-1)$ symbols of the input word, and
proposition  $p_{i}$ is true if, and only if, the $i$-th symbol of the input
word is $+1$.

Let 
$$
  \phi_{(l,C)}  = \lnot B\land \bigwedge\limits_{d=1}^{\lfloor \frac{l}{C+1} \rfloor}\phi_{(l,C,d)}\land \mathsf{frame}_{(l,C)},
$$
where $\mathsf{frame}_{(l,C)}$ is a Boolean formula
encoding that the automaton state is correctly defined, that is,
exactly one proposition 
from each of the sets $\{s_j^{(i,d)}\mid -C\leq j\leq C\}$, for $d=1,\dots, {\lfloor \frac{l}{C+1} \rfloor}$
and $1\leq i \leq \lfloor \frac{l}{d}\rfloor$, is true in every model of $\phi_{(l,C)}$.

The following statement can be easily proved by an investigation of models of $\phi_{(l,C)}$ and
the traces of $\mathcal{A}_{C}$. Notice that although $\phi_{(l,C)}$ encodes 
the traces of $\mathcal{A}_C$ on all subsequences of $\bar{x}$
they all share the same
proposition $B$---as soon as the automaton accepts any of these
subsequences, the entire sequence should be rejected.
\begin{proposition}\label{prop:1}
The formula $\phi_{(l,C)}$ is satisfiable if, and only if, 
there exists a $\pm1$ sequence $\bar x = x_1,\dots, x_l$ of length $l$ of discrepancy $C$.
Moreover, if $\phi_{(l,C)}$ is satisfiable, the sequence $\bar x = x_1,\dots, x_l$ of discrepancy $C$ 
is uniquely identified by the assignment of truth values to propositions $p_1,\dots p_l$.
\end{proposition}

The encoding described above, albeit very natural, is quite wasteful: the size of 
formula $\mathsf{frame}_{(l,C)}$ is quadratic in the number of states.
To reduce the size, in our implementation we use a slightly 
different encoding of the traces of $\mathcal{A}_C$. Namely, we replace
in (\ref{eq:1}) every occurrence of $s_j^{(i,d)}$ with 
a conjunction of propositions representing the numerical value
of $j$ in binary, where the most significant bit encodes the sign of $j$ and
the other bits encode an unsigned  number $0\dots C$ in the usual way. We denote the
resulting formula $\phi^b_{(l,C,d)}$.

For example, for $C=2$ the values $-C\dots C$ can be represented in binary by
$3$ bits. Then $\phi^b_{(l,C,d)}$ contains, for example, 
$$
(\lnot b_2^{(i,d)}\land \lnot b_1^{(i,d)}\land \lnot b_0^{(i,d)})\land p_{id}\rightarrow 
\lnot b_2^{(i+1,d)}\land \lnot b_1^{(i+1,d)}\land  b_0^{(i+1,d)}
$$ encoding the transition from $s_0$ to $s_1$ having read $+1$.

We also exclude by a formula $\mathsf{frame}^b_{(l,C)}$ all combinations of
bits that do not correspond to any states of $\mathcal{A}_C$. For example, for
$C=2$ we have
$$
\begin{array}{ll}
\mathsf{frame}^b_{(l,C)}=\bigwedge\limits_{d=1}^{\lfloor \frac{l}{C+1} \rfloor} \bigwedge\limits_{i=1}^{\lfloor \frac{l}{d}\rfloor +1}
   \bigg[& \lnot (      b_2^{(i,d)}\land \lnot b_1^{(i,d)} \land \lnot b_0^{(i,d)})\land \\
   & \lnot (\lnot b_2^{(i,d)}\land       b_1^{(i,d)} \land       b_0^{(i,d)})\land \\
   & \lnot (      b_2^{(i,d)}\land       b_1^{(i,d)} \land       b_0^{(i,d)})\bigg].
\end{array}
$$
The first conjunct disallows the binary value \texttt{100}, a `negated zero', the other two 
encode that $\mathcal{A}_C$, for $C=2$, does not have neither $s_{3}$ nor $s_{-3}$.
The following statement is a direct consequence of Proposition~\ref{prop:1}.
\begin{proposition}\label{prop:2}
The formula 
  $\phi^b_{(l,C)}  = \lnot B\land \bigwedge\limits_{d=1}^{\lfloor \frac{l}{C+1} \rfloor}\phi^b_{(l,C,d)}\land \mathsf{frame}^b_{(l,C)}$ 
is satisfiable if, and only if, 
there exists a $\pm1$ sequence $\bar x = x_1,\dots, x_l$ of length $l$ of discrepancy $C$.
Moreover, if $\phi_{(l,C)}$ is satisfiable, the sequence $\bar x = x_1,\dots, x_l$ of discrepancy $C$ 
is uniquely identified by the assignment of truth values to propositions $p_1,\dots p_l$.
\end{proposition}

\section{Results}
In our experiments we used the \textsf{Lingeling} SAT solver~\cite{lingeling}
version \textsf{ats}, the winner of the \emph{SAT-UNSAT} category of the SAT'13
competition~\cite{sat13} and the \textsf{Glucose} solver~\cite{Glucose} version
\textsf{3.0}, the winner of the \emph{certified UNSAT} category of the SAT'13
competition~\cite{sat13}.  All experiments were conducted on PCs equipped with
an Intel Core i5-2500K CPU running at 3.30GHz and 16GB of RAM.

By iteratively increasing the length of the sequence,
we establish precisely that the maximal length of a $\pm 1$ sequence of
discrepancy $2$ is $1160$.  On our system it took \textsf{Plingeling}, the
parallel version of the \textsf{Lingeling} solver, about $800$~seconds\footnote{The time taken by the solver varies significantly from experiment to experiment; in one rerun it took the solver just $166.8$ seconds to find a satisfying assignment.} to find
a satisfying assignment.  One of the  sequences of length $1160$ of discrepancy
$2$ can be found in Appendix~\ref{sec:sequence} for reader's amusement.
\begin{proposition}
There exists a sequence  of length $1160$ of discrepancy $2$.
\end{proposition}
When we increased the length of the sequence to $1161$, \textsf{Plingeling} 
reported unsatisfiability. In order to 
corroborate this statement, we also used \textsf{Glucose}.
It took the solver about $21\,500$~seconds to  
compute a Delete Reverse Unit Propagation (DRUP) certificate of unsatisfiability, 
which is a compact representation of the resolution refutation of the given formula~\cite{GoldbergN03}.
The correctness of the unsatisfiability certificate has been independently verified with the 
\textsf{drup-trim} tool~\cite{druptrim}. 
The size of the certificate is about $13$~GB, and 
the time needed to verify the certificate was comparable 
with the time needed to generate it.
Combined with Proposition~\ref{prop:2},  we obtain a computer proof of the following statement.
\begin{theorem}
No sequence of length $1161$ has discrepancy $2$.
\end{theorem}
As there is no finite sequence of discrepancy $2$, there
is no infinite such sequence. So we conclude the following.
\begin{corollary}
The Erd\H{o}s discrepancy conjecture holds true for $C=2$.
\end{corollary}
In an attempt to better understand this result, we looked at the smaller
unsatisfiable subset of $\phi_{(l,C)}^b$ identified by the \textsf{drup-trim}
tool. It turned out that the encoding of some automata traces is not present in
the subset. A further manual minimisation showed that, although $\lfloor
\frac{1161}{3}\rfloor$ is $387$, to show unsatisfiability  it suffices to consider subsequences of 
$x_1,\dots, x_{1161}$ of the form $x_d,\dots, x_{kd}$ for the values of $d$ ranging
from $1$ to $358$. It remains to be seen whether or not this observation can be helpful for 
a human proof of the conjecture

We also applied our methodology to identify sequences of discrepancy $3$, however,
we did not manage to prove the conjecture.
Having spent 3~days, 7~hours and 30~minutes
(or $286247.9$~seconds total),
on the encoding of the problem using 356\,048~variables and 4\,342\,612 clauses
\textsf{Plingeling} has successfully 
identified a sequence of length $13\,000$ with discrepancy $3$. 
The encoding and the generated
sequence can be found in~\cite{KL14}.
\begin{proposition}
There exists a sequence  of length $13\,000$ of discrepancy $3$.
\end{proposition}
Unfortunately, our attempts to improve this result did not succeed: 
\textsf{Plingeling} did not return any answer  on the encodings 
of discrepancy $3$ problems for sequences of 
length $14\,000$ and $16\,000$
even within $1\,550\,000$ and $2\,280\,000$~seconds, respectively; the
\nb{AL: replaced 13000 with 1300 000}
computations are still going on and the problem is still open.

\section{Discussion}
We have demonstrated that SAT-based methods can be used to tackle the
longstanding mathematical question on the discrepancy of $\pm 1$ sequences. 
For EDP with $C=2$ we have identified the exact boundary between satisfiability
and unsatisfiability, that is, we found the longest discrepancy $2$ sequence
and proved that no larger sequence of discrepancy $2$ exists.
There is, however, a noticeable asymmetry between these findings.  The fact
that a sequence of length 1160 has discrepancy $2$ can be easily checked either
by a straightforward computer program or even manually.  The negative witness,
that is, the DRUP unsatisfiability certificate, is probably one of longest
proofs of a non-trivial  mathematical result ever produced.  Its gigantic size
is comparable, for example, with the size of the whole Wikipedia, so one may
have doubts about to which degree this can be accepted as a proof of a
mathematical statement. 

But this is the best we can get for the moment. Essentially, the unsatisfiability proof corresponds to
the verification that the search in a huge search space has been done
correctly and completed without finding a satisfying assignment.  It is a
challenging problem to produce a compact proof more amenable for human
comprehension.

Finally notice that apart from the obtained results the proposed methodology can be
used to further experimentally explore variants of the Erd\H{o}s problem as well as
more general discrepancy theory problems.  
%
%

%
%
%
%
\appendix
\section{One of the sequences of length $1160$ having discrepancy $2$}\label{sec:sequence}
We give a graphical representation of one of the sequences of length $1160$ 
obtained from the satisfying assignment computed with the \textsf{Plingeling} solver.
Here $+$ stands for $+1$ and $-$ for $-1$, respectively.

\noindent{}\tt 
- + + - + - - + + - + + - + - - + - - + + - + - - + - - +\\
+ - + - - + + - + + - + - + + - - + + - + - - - + - + + - \\
+ - - + - - + + + + - - + - - + + - + - - + + - + + - - - \\
- + + - + + - + - + + - - + + - + - + - - - + + - + - - +\\
+ - + + - + - - + + - + - - + - - - + - + + - + - - + + - \\
+ + - + - - + - - + + - + + - + - - + + - + - - + + + - +\\
- + - - - - + + + - + - - + - - + + + - - - + + - + + - + \\
- - + - - + + + - - + - + - + - - + - + + + - + + - + - - \\
+ - - + + - + - - + + - + + - + - - + - - + + - - + + + -\\
- - + + + - + - - - + + - + - - + + - - + - + - - + - + + \\
+ - + - - + + - + + - + - - + + - + - - + - - + + - + - -\\
+ + - - + - + + - + - + - - + - + - + + - + - - + + - + -\\
- + - - + + - + - + - + + - + - + - + + - - - + - + - - + \\
+ + + - - + - - - + + - + - + + - + - - + + - + - - + - -\\
+ + - + - - + + + + - - + - - - + - + + + + - - + - - + +\\
- + + - + - - + + - + - - + - - + + - + - - + + - + + - +\\
- - + + - + - - + - - + + - + + - + - - - - + + + - + - - \\
+ + - - + + + - - - + - + + - + - - + - + + - - - + - + +\\
- + + - + - - + - - + + - - + + + + - + - - + - - + - - + \\
+ + + - - + - - + + + - - - + + - + + - + - - + + - - + - \\
+ - - + - - + + - + + - + - - + - - + - + + + - + + - + - \\
- + - - + + - - + - + + - + + - + - - + - - + - - + + - + \\
+ - + - - + + + - - - + + - + - - + + - + + - - - + + + - \\
- - + + - + + - - - - + + + - - + - + + - + - - + - - + +\\
- + - - + + - + + - + - + + - - + + - - + + - - - - + + + \\
- + + - - + + - - - - + + - + + + - - + + - - - + + + - -\\
- - + - + - + + - + + - + + - + - + - - - - + + + - - + +\\
- + - - + + - + + - + - - + - - + - - + + - + - - + + - + \\
+ - + - - + + - - + - + - - + - + - + - + + + + - - - + -\\
+ - + + - - + - - + - + - + - + + - + - + + + - - + - + - \\
- + - - + - + + + - - + - + + + - - - + + - + - - + - - +\\
+ - + + - - + + - - - + + - + - + + - - + + - + - - - + - \\
+ + - + - - + - + + - - + + - + - - + + - + - - + - + + + \\
- + - - + + - - + - + - + + + - - + - + - - + + - + + - +\\
- - + - - + - + + - - - + - + + - + - + + - - + + - + - -\\
+ + + - + - - - - + + - - + - + + - + - + + - - + + - + - \\
- + + - + - + + - - + + - + - - - + - + + - + - - + + + - \\
- - - + - + - + + - - + + - + - - + + - + + - + + - + - - \\
+ - - + - - + + + + - - - + + - - - + - + - + + - + - + + \\
+ - - + - + + - - + - + - - + - + - + + - - - + + + - + + \\
\end{document}